\documentclass[sigconf]{acmart}
\makeatletter
\renewcommand\@formatdoi[1]{\ignorespaces}
\makeatother
\setcopyright{none}
\renewcommand\footnotetextcopyrightpermission[1]{}
\AtBeginDocument{%
  \providecommand\BibTeX{{%
    \normalfont B\kern-0.5em{\scshape i\kern-0.25em b}\kern-0.8em\TeX}}}

\usepackage{dirtytalk}

\begin{document}

\title{Designing a Secure Device-to-Device File Transfer Mechanism}

\author{Chaitanya Rahalkar}

\email{cr@gatech.edu}
\authornotemark[1]
\orcid{0000-0003-2350-9793}
\affiliation{%
  \institution{Georgia Institute of Technology}
  \streetaddress{North Ave NW}
  \city{Atlanta}
  \state{Georgia}
  \country{USA}
  \postcode{30332}
}

\author{Anushka Virgaonkar}
\authornote{Both authors contributed equally to this research.}

\email{avirgaonkar3@gatech.edu}
\affiliation{%
  \institution{Georgia Institute of Technology}
  \streetaddress{North Ave NW}
    \city{Atlanta}
    \state{Georgia}
  \country{USA}
  \postcode{30332}
}

\begin{abstract}
  Secure, reliable, and fast transfer of files across the Internet is a problem attempted to be solved through many application-layer protocols. In this paper, we aim to design a secure, reliable, open-design, and performant file transfer protocol that is inspired by the WebRTC protocol stack. Traditionally, transferring files involves a publicly exposed (available on the public network) third-party server that serves the uploaded files to the receiver. Here, the third-party server has to bear the storage and bandwidth cost to transfer the files between the two parties. We propose a protocol that uses a relay server to relay the files from the client to the server. A relay server has several advantages over a regular file-hosting server. Firstly, a relay server does not retain the uploaded files, it simply relays them. Secondly, a relay server has a full-duplex communication channel and therefore the receiver is not required to wait for the sender to upload the files completely. In this paper, we study available file transfer approaches and their known flaws. We propose our idea and compare our stack with the WebRTC stack. Finally, we perform empirical analysis and, benchmark our device-to-device transfer approach along with other available options including WebRTC. 
\end{abstract}   



\maketitle
\pagestyle{plain}
\section{Introduction}

Transferring files over the internet may seem like a simple task and many protocols are available to achieve it. However, keeping the transfer secure, fast, open, and free is not achievable by most of the protocols. They struggle in achieving all of these goals at once. Ideally, in file transfer, we want to provide security which means we want to ensure the confidentiality and integrity of the file data in the transfer. The privacy of the user data must be protected as well. Often, when relying on third-party services, the security and privacy of user’s documents are debated. In 2012, Dropbox, a cloud storage platform, suffered a cyberattack that leaked confidential data of about 68 million users\cite{Dropbox_hack}. Placing trust in such services is becoming difficult not only because of the devastating cyberattacks that they suffer but also because of their concerning privacy policies and the inherent absence of transparency. Cloud-based file transfer/storage implementations rarely provide transparency about the encryption schemes and storage design mechanisms which violates the \say{Open Design} security principle. There have been cloud platforms like Mega, which do provide some implementation details in their security whitepaper. However, some overhead issues have been found in their client-side encryption scheme \cite{henziger2019overhead}. Many of the services offer a limited amount of free storage, after which the user has to choose a paid plan to continue to use the service. \\

In this paper, we aim to focus mainly on achieving three goals: security, performance, and transparency. We also aim to provide an empirical analysis of the security of our implementation in comparison to the many existing file transfer approaches. Our relay-server-based approach is highly inspired by the WebRTC protocol stack. Even though the WebRTC protocol is not designed for arbitrary data communication, we have managed to replicate this with some additional changes that ensure that the protocol is devoted entirely to transferring files. The relay-server-based approach poses a significant disadvantage from the costing perspective because it requires the relay server to be hosted on a cloud provider. Apart from the VM configuration, cloud storage services also charge based on the amount of network bandwidth used.  Typically, the relay-based approach would be costly to bear if large files are to be transferred. We study some trade-offs of the proposed approach and the WebRTC-based file transfer approach. We attempt to make inferences about which approach would be ideal given the constraints.  With the empirical results, we aim to verify whether our implementation meets the required goals. \\

\section{Background}
Many traditional file transfer approaches have been proposed, but very few of them have managed to satisfy the three primary goals - security, transparency, and performance. 
Consider a scenario where two parties: sender and receiver, have agreed upon the sharing of a particular file. Commonly, HTTP would be used to transfer files between them on the Internet. However, HTTP was traditionally designed to be a text protocol and was not optimized initially, to send large binary blobs. However, with the recent developments in the HTTP protocol, sending binary blobs and binary data packets is now equally optimal as compared to the other competitors. \\

The File Transfer Protocol (FTP) was introduced in 1971, specifically, for transferring files. However, security is a big concern in the case of this half-a-decade-old protocol \cite{Postel_Reynolds}.  FTP was not designed with security in mind. Even though some security improvements have been proposed for FTP, like tunneling the FTP protocol over SSH (known as SFTP), they still have had a lot of security issues and vulnerabilities over the years \cite{ylonen2006secure}\cite{openssh}. \\

Protocols like SCP, which were designed with security in mind, however, require port forwarding when the sender and receiving parties are behind a NAT. Although the file transfer is still achievable, it is not simple. The protocol observes a lot of overhead in terms of mandatory acknowledgments that when failed, call for re-transmission. Moreover, SCP is outdated, inflexible, and contains several vulnerabilities.\\

Transferring files usually involves a third, trusted entity to host the uploaded file. Introducing such a trusted entity is always a big overhead since the file transfer becomes a two-stage process - 
\begin{enumerate}
    \item Sender uploads the file to the server, and
    \item Receiver downloads the file from the server.
\end{enumerate}

They have file size limits and storage limits. Due to performance issues, lack of transparency, and privacy concerns, this approach does not meet all of our goals. \\

Direct device-to-device file transfer can be challenging due to NAT. One or both of the clients can be behind a NAT network, which makes it difficult to correctly route the packets to the intended device. Even though the packet may reach the NAT router correctly, the NAT router will often drop these packets or won't transfer them to the intended recipient. However, if both of the devices are exposed to the public network, the transfer agreement and process becomes considerably simpler. In such a case, the requirement of a relay is eliminated. However, a third-party server is required only for the initial message exchange scheme that connects the two clients. Peer-to-peer connection for two devices behind a NAT involves a technique called \say{NAT traversal}. This is done through UDP/TCP hole-punching, STUN and TURN servers, etc. There has been a proposal that attempted to traverse the NAT using fake ICMP messages. However, this proposal was a \say{hack}, that deceived NAT routers and does not work with all NAT routers \cite{muller2010autonomous}.\\

There have been some device-to-device file transfer approaches like the Magic Wormhole \cite{williams2019magic}, which also uses a relay server to relay files between the sender and receiver. Our relay-based protocol implementation counters some of the limitations observed in this approach. 

\section{Approach}
Our relay-based device-to-device file transfer implementation involves several components that successfully ensure that the transfer is handled securely and reliably. These components collectively aim to establish our goal. 

\subsection{Components}
In our proposed idea, we have the following four main components: 
\begin{itemize}
    \item Relay Server
    \item Password Authenticated Key Exchange
    \item IP Exchange Scheme
    \item Device Clients
\end{itemize}

\subsubsection{Relay Server}
We propose an end-to-end encrypted file transfer approach to transfer files between two computers. This approach does not require the file to be uploaded to a third-party server, and the receiver is not required to wait for the sender to upload the file completely. We make use of a relay server that establishes full-duplex communication with the sender and receiver. This full-duplex communication is achieved using WebSockets. The connection is secured using TLS (WSS://). The relay server simply relays the data packets from the sender to the receiver. The data packets are small chunks that the entire file data is divided into. In this whole transfer process, no data is stored on the relay server. The relay server does the job of \say{gluing} the two incoming TCP connections and transfers data. This relay server is similar to what a TURN (Traversal Using Relays Around NAT) does in the WebRTC stack. Initially, we check if both the clients (sender and the receiver) are publicly exposed by sending direct packets directed to each other (after the IP address exchange). If the connection is successful, then a truly device-to-device file transfer mechanism can be achieved. However, if the connection fails (either due to firewall restrictions or one or both of the clients being behind a NAT), only then the relay server is involved. However, in a typical scenario, one or more of the clients are behind a NAT router/firewall. Devices or machines that are exposed to the public network are mostly servers hosting some publicly available content or service. \\

The communication in the mechanism is made completely end-to-end encrypted so even the relay server cannot read the contents of the data packets. If the relay server is malicious and intends to attack the integrity of the data packets, such an attempt would be detected by the receiver, and the receiver can then discard those data packets. \\

During our initial benchmarks, we had observed that the bandwidth charges for the cloud-hosted relay were very high. This is because data centers charge for ingress and egress bandwidth consumption. Ingress bandwidth consumption is a result of the sender sending the data packets to the relay, and the egress bandwidth results from the relay sending those packets to the receiver. As a countermeasure, the WebRTC-based transfer approach is explored, since it is a truly device-to-device file transfer approach in almost all scenarios. It has been elaborated in the further subsections of the paper. \\

\subsubsection{Password Authenticated Key Exchange}

For securing the end-to-end transfer of the file by establishing a secure channel, we implement a password-authenticated key exchange. It is a method in which a sender and a receiver agree upon cryptographic keys that are derived from a secret passphrase that is known to one or both of them. In the file transfer, the data packets that are transmitted between the sender and the receiver are encrypted using a shared session key which is established from the secret passphrase. For each session, a new session key is generated and used. The protocol secures the system even if both parties choose weak passwords. This increases the usability of our design without having to compromise on security\cite{SPAKE2}.\\

\begin{figure}[ht]
  \centering
  \includegraphics[width=\linewidth]{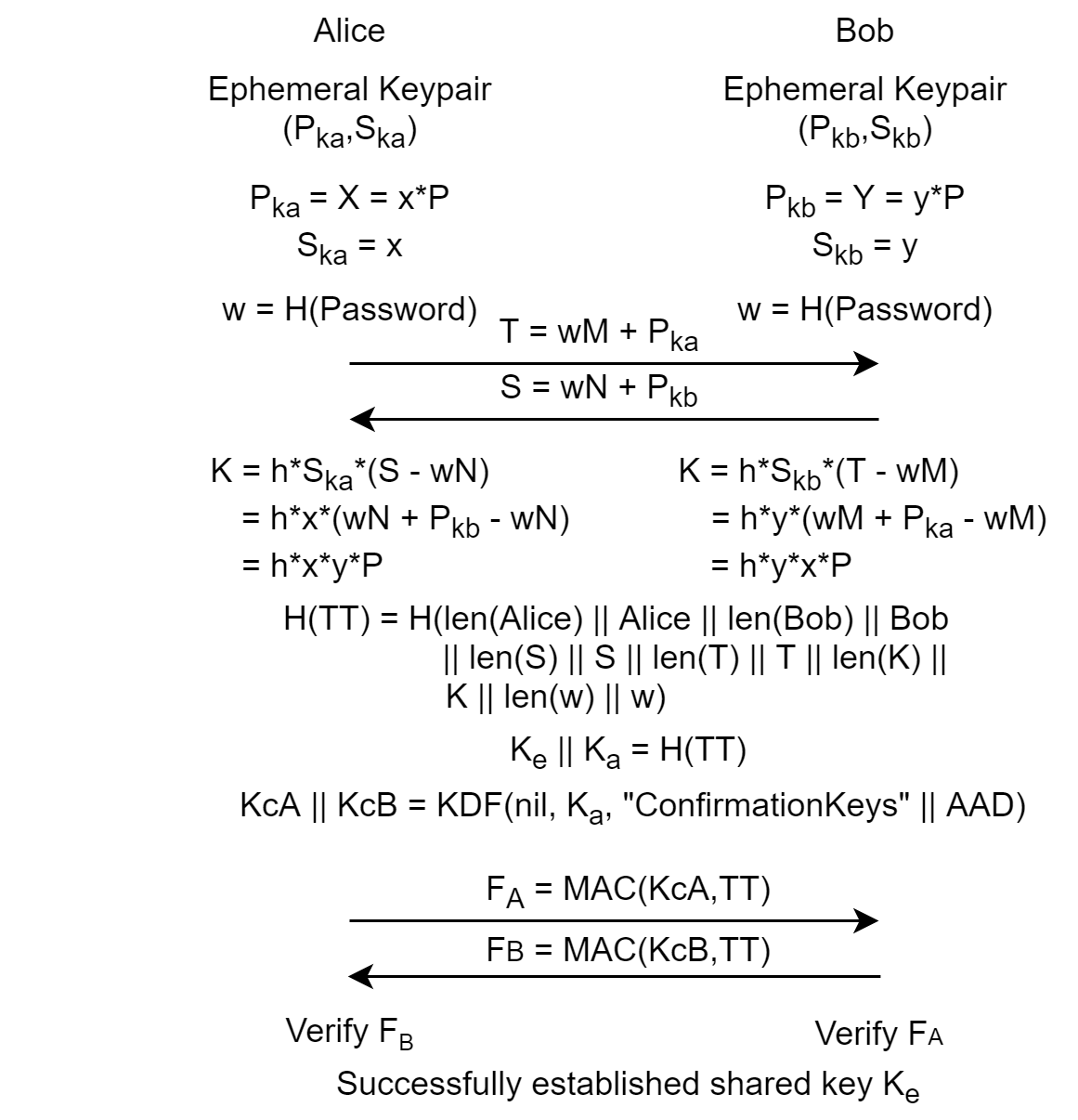}
  \caption{The sPAKE2 Protocol}
  \label{fig:spake}
\end{figure}

We have used sPAKE2, which is a symmetric form of Password Authenticated Key Exchange (PAKE), in our protocol. Consider that Alice and Bob want to derive a shared secret key securely. They both know a password. Fig. \ref{fig:spake} shows the message exchange between Alice and Bob that leads to the generation of the shared key. The following are the steps that would be followed in the sPAKE2 protocol:  \\
\begin{enumerate}
    \item $G$ is an elliptic curve group of the order $p*h$ where $p$ is a large prime and $h$ is the cofactor. The computational Diffie-Hellman (CDH) assumption holds in $G$. $M$ and $N$ are in the subgroup of G, and P is the generator of the subgroup of G. Alice and Bob agree on these parameters.
    \item Alice generates an ephemeral public key \begin{align*}
    X &= x*P,
    \end{align*}
    where $x$ is the private key $S_{kb}$ which is a randomly chosen integer that lies in $[0,p)$. Alice uses a memory-hard hash function to compute the hash of the password, which is $w$,
    \begin{align*}
    w&=H(password),
    \end{align*}
    and calculates 
    \begin{align*}
    T&=w*M+X,
    \end{align*}
    which becomes her public share. Similarly, Bob generates his ephemeral public key
    \begin{align*}
    Y&=y*P,
    \end{align*}
    where $y$ is his private key $S_{ka}$ which is a randomly chosen integer that lies in $[0,p)$ and calculates
    \begin{align*}
    S&=w*N+Y,
    \end{align*} which is his public share.
    \item Alice and Bob share their public shares with each other.
    \item Both parties calculate a shared group element K, with the information that they have received from each other. Alice calculates $K$ as:
    \begin{align*}
    K&=h*S_{ka}*(S-w*N)
    \end{align*}
    and Bob calculates it as 
    \begin{align*}
    K&=h*S_{kb}*(T-w*M)
    \end{align*}
    We can show that they both calculate the same value because 
    \begin{align*}
    h*S_{ka}*(S-w*N) &= h*x*(w*N+Y-w*N)\\
    &= h*x*y*P.\\
    h*S_{kb}*(T-w*M) &= h*y*(w*M+X-w*M)\\
    &= h*y*x*P 
    \end{align*}
    So, they both calculate the same value of $K$ which is $h*x*y*P$. 
    \item However, $K$ is not used as the shared secret because it is not secure against man-in-the-middle attacks. Alice and Bob calculate 3 shared secrets: $K_{e}, KcA, KcB$. The hash of the transcript $TT$ is computed as follows:
    \begin{align*}
    \begin{split}
    H(TT)&=H(len(Alice) || Alice || len(Bob) || Bob
    || len(S) ||\\&S || len(T) ||T || len(K) ||K || len(w)||w),  
    \end{split}
    \end{align*}
    where Alice and Bob are their identifiers.
    The symmetric secrets $K_{e}$ and $K_{a}$ are calculated as follows:
    \begin{align*}
    K_{e} || K_{a} &= H(TT),\\
    |K_{e}| &= |K_{a}|
    \end{align*}
    The keys $KcA, KcB$ are calculated as follows:
    \begin{align*}
    KcA || KcB &= KDF(nil, K_{a}, "ConfirmationKeys" || AAD)
    \end{align*}
    \begin{align*}
    |KcA| &= |KcB|,
    \end{align*}
    where KDF is a key-derivation function that derives a key by taking $K_{a}$, a string "ConfirmationKeys", and Additional Authenticated Data (AAD) (or nil if it is not provided) as input. AAD is the associated data that Alice and Bob share which is separate from their identities that they may want to include in the protocol execution. One example of AAD is a list of supported protocol versions if sPAKE2(+) were used in a higher-level protocol which negotiates use of a particular PAKE\cite{Ladd_2020}.
    \item To verify if both the parties were able to successfully generate the secret, the final step which is the key verification confirmation step is carried out. Alice computes a message authentication code (MAC) $F_A$ of the transcript $TT$, with $KcA$ as the authentication key for the MAC as follows:
    \begin{align*}
    F_A &= MAC(KcA,TT) 
    \end{align*}
    Alice sends $F_A$ to Bob. Similarly, Bob calculates a MAC of the transcript with $KcB$ as the key as follows:
    \begin{align*}
    F_B &= MAC(KcB,TT) 
    \end{align*}
    Bob sends $F_B$ to Alice. Both parties verify the MACs that they receive by calculating $F_A$ or $F_B$ on their end. If the key confirmation verification is successful, then both parties know that have successfully derived the shared key.
\end{enumerate}

\subsubsection{IP Exchange Scheme}

The relay server must have the sender’s public IP address and the receiver’s public IP address. One or more of the parties can be behind a NAT router. Therefore, to connect to these peers and establish a relay scheme, the relay server must have the public IP addresses of both of the clients. This is important for the initial sPAKE2 handshake that authenticates both the clients and authorizes the sender to send the data to the receiver.

\subsubsection{Device Clients}
The device client is the end node in our file transfer i.e the sender or the receiver. It uses TCP sockets for reliable transport. The sPAKE2-based key authentication protocol is used for secret sharing, which is a type of a balance password-authenticated key exchange that uses the same password for negotiating and authenticating a shared key. The file transfer using this end-to-end secure channel works even if the sender and receiver are behind different routers. Although, having a hardened firewall or restrictive NAT may not allow establishing such a channel.

\begin{figure}[ht]
  \centering
  \includegraphics[width=\linewidth]{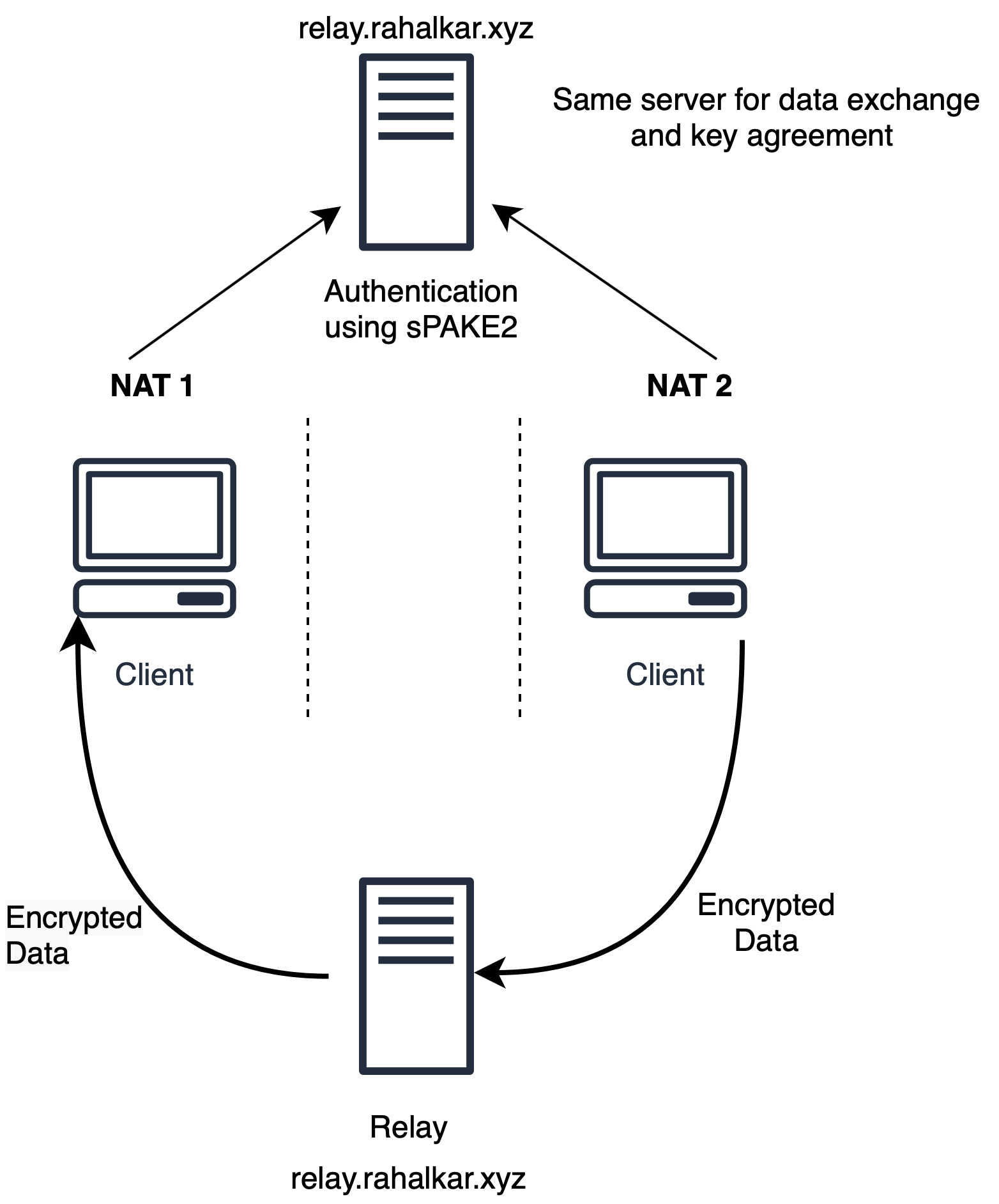}
  \caption{End-to-End Flow of the Transfer Mechanism}
  \label{fig:all}
\end{figure}

When the sender and receiver want to exchange a file, they communicate with each other to share a secure shared key. The sPAKE2-based key authentication protocol is used for the same. The sender divides the file into data packets of 16384 bytes and encrypts them. The receiver puts in this passphrase and a secure data transfer channel is established and the sender sends the encrypted data packets. The relay server obtains the receiver's address from the IP exchange scheme and relays the packets to the receiver. Any modifications in the data packets by the relay would be detected by the receiver because the receiver performs integrity checks and verifies the authenticity of the sender. The receiver enters the password and the file is reconstructed from the data packets. Fig. \ref{fig:all} shows the entire protocol implementation flow. 
\\

\subsection{Implementation}

Before any of the transfers happen, a relay is hosted on a publicly available network so that both the sender and the receiver can communicate with it. It uses a minimal SQLite3 database to store connection-related information (passphrase, IP addresses, etc.) In our benchmarks, the relay server was hosted on \href{https://relay.rahalkar.xyz}{relay.rahalkar.xyz}. The mechanism has been implemented using Golang and is usable as a CLI tool. The server specifies the file to be sent to the CLI program. The file is broken into chunks of 16K Bytes. A random passphrase is generated on the sender's side. The sender now communicates with the relay server and tries to initiate the sPAKE2 handshake. The secret passphrase should now be communicated through a secure channel (End-to-end encrypted call or message service can be used) to the receiver before the communication can continue. When the receiver attempts to receive the file, the CLI program asks for the secret passphrase. Once the receiver enters the secret passphrase, it's sent to the relay server. The relay server links the two clients based upon the secret passphrase. This relay server is responsible for exchanging the sPAKE2 messages and establishing a shared key between the two clients. Once the two clients are successfully authenticated, and the sender is authorized to send the files, the binary chunks are encrypted using NaCl's authenticated encryption scheme (known as Crypto Secret box) \cite{bernstein2009cryptography}. Authenticated encryption preserves message integrity and confidentiality. It protects against chosen-ciphertext attacks. As and when the encrypted packets are sent to the relay server, they are immediately relayed to the receiver. Finally, when all the chunks are received, the receiver reconstructs the file after verifying the authenticity and integrity. 

\subsection{Security of Protocol}

The file transfer protocol is secure against malicious relay servers. If the relay server attempts to modify any data packets, such an action would be detected by the receiver because of the use of authenticated encryption. The security of the sPAKE2 protocol depends on the computational Diffie-Hellman problem. Using sPAKE2, a weaker password can still provide strong security. The use of a memory-hard hash function to generate the hashes of various values in the protocol helps slow down an attacker. Man-in-the-middle attacks are not possible in this protocol. Consider an eavesdropper Eve, who pretends to be one of the parties and tries to guess the password possessed by Alice. Eve and Alice would share their public shares and would calculate K, where K would be different for both of them, considering that Eve does not guess the right password. When Eve sends her MAC to Alice, the verification would fail. Alice can blacklist Eve to lock her out. Offline analysis of the password or any other parameters is not possible because the protocol relies upon online communication between the parties. A new key is used for a new session. Due to all these factors, sPAKE2 is a secure cryptographic protocol that is successful in providing strong security even with weak passwords.

\subsection{Web Real-Time Communication}
Another protocol that meets our goals of security, performance, transparency, and cost-effectiveness is Web Real-Time Communication (WebRTC). The relay server that we use in our protocol has an added cost to it, which is eliminated in WebRTC. WebRTC provides real-time file-sharing capabilities without having to store the file on a server. Even though the communication is device-to-device, a signaling server is used to establish the connection (handshake and exchange of several parameters) between the two devices. WebRTC was typically designed for real-time communication channels like VoIP, and low-latency applications like streaming, etc. However, the WebRTC stack has the \say{RTCDataChannel} API that allows us to send arbitrary data packets \cite{loreto2014real}. The WebRTC API was designed to work with existing browsers and was implemented for Javascript. However, several ports of the API into other programming languages (E.g. Aiortc in Python) have allowed us to use the API without depending on a browser.  \\

The WebRTC data transfer happens through two important APIs - 
\begin{enumerate}
    \item RTCPeerConnection - This API is used to initiate and complete the WebRTC connection. When the sender wants to send data to the receiver,  an SDP offer (Session Description Protocol) is generated by the sender as a part of the connection initiation scheme. This SDP is to be sent to the receiver (sent using a third-party signaling server), and the receiver responds with its own SDP answer (sent using the same signaling server).  With this exchange, the two clients establish a connection between them. 
    \item RTCDataChannel - This API is used to exchange arbitrary data between the two devices. The RTCDataChannel API is similar to Websockets with some changes in the data delivery properties \cite{vogt2013leveraging}. 
\end{enumerate} 

\begin{figure}[ht]
  \centering
  \includegraphics[width=\linewidth]{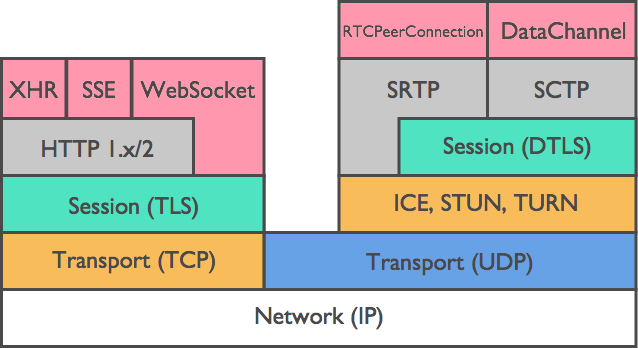}
  \caption{WebRTC Protocol Stack (From \href{https://webrtc-security.github.io}{webrtc-security.github.io})}
  \label{fig:webrtc}
\end{figure}

Whenever both parties are behind a NAT router/firewall, getting the public IP addresses of the parties can be difficult. Therefore, a STUN server is used to ensure both parties know each others' IP addresses. 
WebRTC is secured using DTLS (Datagram Transport Layer Security) \cite{rescorla2013webrtc}. DTLS is based on the TLS protocol that is used to secure transport layer packets.\\ 
The WebRTC stack is shown in fig. \ref{fig:webrtc}. The figure indicates the UDP-based WebRTC stack and its counterpart in the TCP stack using the same colors.  \\

WebRTC, however, has problems working through heavily restricted firewalls. It also requires a browser that supports WebRTC to initiate and complete the transfer. Moreover, the communication is over UDP, which is not reliable. At times, if the device-to-device communication is not achieved, WebRTC proposes a fallback option using a TURN server. This model is similar to our relay-based approach as proposed earlier. However, due to the wide acceptance of WebRTC, only in cases where firewalls are heavily restricted or WebRTC is disabled, the fallback option would be used. Hence, in most cases, the transfer would be truly device-to-device. The significant advantage that WebRTC has over our relay-based implementation is that the TURN server is only used as a fallback option when the device-to-device connection fails. At all other times, the transfer is done without the involvement of a third party. AioRTC, a Python-based WebRTC implementation library was used to test and benchmark the WebRTC-based transfer process.  

\subsection{Comparing WebRTC and our Relay-Based Approach}
Table \ref{tab:comparison} shows a direct comparison between the components used by WebRTC and our relay-based transfer approach. WebRTC requires several extra components like a STUN, TURN, and Signalling server to complete the stack. All the functionalities implemented by these servers are handled by our single relay server. Therefore, in our approach, we're able to handle the complete transfer mechanism simply by deploying the relay server on a publicly available computer. The only significant advantage that WebRTC poses is that it relies on the TURN relay during fallback. 

\begin{table}[hbt!]
\begin{tabular}{|p{0.23\linewidth}|p{0.25\linewidth}|p{0.4\linewidth}|}
\hline
Comparison Factor           & WebRTC                    & Relay-Based Transfer                                                                                                         \\ \hline
Security                    & DTLS                      & sPAKE2 + NaCl Crypto                                                                                                         \\ \hline
Relaying Data               & TURN                      & Handled by the relay server                                                                                                  \\ \hline
Requirement of Relay Server & Only as a fallback option & Mandatory for all cases except  when both devices are exposed to public network \\ \hline
Connection Establishment    & Signalling server         & Handled by the relay server                                                                                                  \\ \hline
IP Exchange                 & STUN server               & Handled by the relay server                                                                                                  \\ \hline
\end{tabular}
\caption{Comparison between WebRTC and our relay-based file transfer approach}
\label{tab:comparison}
\end{table}

\begin{table*}[hbt!]
\begin{center}
\begin{tabular}{|p{0.1\linewidth}|p{0.15\linewidth}|p{0.15\linewidth}|p{0.15\linewidth}|p{0.15\linewidth}|p{0.15\linewidth}|}
\hline
 &
  Overall Security &
  End-to-End Encryption &
  Source &
  Intermediary Involved &
  Eavesdroppers \\ \hline
Google Drive &
  Account-based &
  No, keys held by Google &
  Closed Source &
  Yes (Google Servers) &
  Google, CAs \\ \hline
WebRTC &
  DTLS &
  Yes &
  Open protocol design, FOSS libs &
  No (STUN Server to get Public IP) &
  None. Peer-to-peer design \\ \hline
Our Implementation &
  sPAKE, NaCl Crypto &
  Yes &
  Open protocol design &
  Yes (Relay server) &
  None, property of PAKE \\ \hline
FTP &
  Password-based access to servers &
  No &
  Open protocol design, FOSS libs &
  Yes (FTP server) &
  FTP server host \\ \hline
Email &
  TLS &
  Only in PGP/S-MIME &
  Open protocol design, FOSS libs &
  Yes (Mail server) &
  Mail servers (None in PGP/S-MIME) \\ \hline
SCP &
  AES, RSA/ ECC, HMAC &
  Yes &
  Open protocol design, FOSS libs &
  No &
  None \\ \hline 
\end{tabular}
\caption{Empirical Results}
\label{tab:empirical}
\end{center}
\end{table*}

\begin{table*}[hbt!]
\begin{center}
\begin{tabular}{|p{0.15\linewidth}|p{0.1\linewidth}|p{0.1\linewidth}|p{0.1\linewidth}|p{0.1\linewidth}|}
\hline
  &
  1MB &
  100MB &
  512MB &
  1GB \\ \hline
  SCP & 
  3.28s &
  49.69s &
  4m 29s &
  9m 13s\\ \hline
  Our Implementation &
  3s (receiving) &
  49s (receiving) &
  3m 12s (receiving) &
  6m 28s (receiving)\\ \hline
  WebRTC &
  1.80s &
  41.32s &
  4m 13s &
  7m 2s \\ \hline
  FTP &
  0.37s &
  55s &
  7m 12s & 
  16m 32s \\ \hline

\end{tabular}
\caption{Benchmarking Results}
\label{tab:benchmark}
\end{center}
\end{table*}

\section{Results}
In this paper, we have also established empirical analysis and performance benchmark results. We have also provided a comparative analysis between our approach and the different file transfer implementations available. We have used 5 parameters that take into account the security of these implementations. These parameters talk about-
\begin{itemize}
    \item the overall security (channel + data),
    \item usage of end-to-end encryption,
    \item whether the protocol or program is open-sourced,
    \item whether an intermediary is involved and,
    \item whether there can be possible eavesdroppers in the communication medium. 
\end{itemize}

In table \ref{tab:empirical}, our two experiments (relay-based + WebRTC-based) are compared with other approaches like FTP, SCP, etc. Even though SCP has no flaws as per this table, it was declared outdated by OpenSSH due to multiple vulnerabilities found in the protocol. Also, it does not have a regulated access control mechanism. An exploit in SCP can be used to move laterally and affect other SSH-based implementations. Our implementations using the relay server and WebRTC are considerably secure due to the usage of modern, secure, robust, and open-design protocols. \\

As shown in table  \ref{tab:benchmark}, the performance benchmarks were taken for binary files of size 1MB, 100MB, 512MB, and 1GB. We have used SCP, our relay-based implementation, our WebRTC DataChannel API-based implementation, and FTP to test out performance benchmarks. The performance of our relay-based implementation also depends on the location of the relay server. This performance could have improved marginally if the relay server was at a location close to both of the clients. From the benchmarks, we were able to conclude that for smaller file sizes, the performance of WebRTC and our relay-based implementation was similar. However, our relay-based approach had a significant advantage when the file sizes were 512MB and 1GB. Even though SCP managed to show close to similar results, there are several complications in the protocol that we talked about previously. Even though WebRTC looked performant there were some hiccups and limitations we observed that are discussed later in the paper. \\

The relay server was hosted at a Google Cloud VPS at a data center in Iowa and a DigitalOcean VPS in Oregon was used for the other intermediary-involved tests. Tests for all other implementations were conducted in a similar environment. Even though these tests depend on the kind of environment used, the internet speed, etc., using a similar environment for all the tests allowed us to gauge the performance. \\

\section{Discussion}

With our implementation of a secure end-to-end encrypted file transfer protocol, we have achieved our main goals of security, performance, and transparency. We observed that other file transfer and storage services like Google Drive, Dropbox, and protocols like FTP and SCP, do not meet our main goals. \\

There were some problems observed during the transfer of files larger than 1GB in-case of WebRTC based transfers. WebRTC data channels are not recommended for large file transfers since the data channel packet size is capped at 16KB. Also, it is a UDP-based protocol and therefore can introduce unreliability. Our relay-based implementation can send big files efficiently and the communication is using TCP, which is reliable as it provides delivery guarantees. However, using the relay server can be costly. \\

Our protocol is designed for providing security to users who want to share files easily, securely, and privately. However, issues such as piracy and censorship can arise. As with any other end-to-end encrypted file transfer protocol, our protocol fails to control the transfer of censored or pirated material. Law enforcement agencies would not be able to observe the conduction of such activities.

Our emphasis on security and privacy can make our protocol useful for an organization that does not want to rely on third-party services for file transfer. The organization can invest in relay servers and ensure that their files which are confidential in nature are transferred securely. Further enhancements to security and privacy could be achieved by hosting the relay server on the Tor network. But this would significantly impact the transfer performance for larger files. In this approach, issues in performance may be observed, but given the high assurance of security and privacy, our protocol is worth the consideration. \\

\section{Conclusion}

In this paper, we have successfully designed and developed a secure end-to-end encrypted file transfer protocol that satisfies the goals of security, performance, and transparency. We compared our approach with the many existing file transfer approaches and have presented our results. Few use cases were discussed as well.

\bibliographystyle{unsrt}
\bibliography{sample-xelatex}

\end{document}